\documentclass[11pt]{article}
\usepackage{moriond,epsfig}

\def\be{\begin{equation}}
\def\ee{\end{equation}}
\def\bea{\begin{eqnarray}}
\def\eea{\end{eqnarray}}

\def\galm{\mathrm{{gal}}}
\def\cmbm{\mathrm{{cmb}}}

\def\comic{CMB }

\def\Cross{{\cal X}}

\def\ee{\mathrm{e}}
\def\pabo{{\nabla }}
\def\bo{{\Delta }}

\def\CosVar{{\mathrm CosVar}}

\begin{document}
\vspace*{4cm}
\title{CROSS-CORRELATING CMB POLARIZATION WITH LOCAL LARGE SCALE STRUCTURES}

\author{K. BENABED\protect\(^{\dag}\protect\), F. BERNARDEAU\protect\(^{\dag}\protect\), L. VAN WAERBEKE\protect\(^{\ddag}\protect\) }

\address{\protect\(^{\dag}\protect\)SPhT, C.E. De Saclay, 91191 Gif-Sur-Yvette, France\\
\protect\(^{\ddag}\protect\)CITA, 60 St George Str.,Toronto M5S 3H8 Ontario, Canada}
\maketitle
\abstracts{
We study heterogeneous quantities that efficiently cross-correlate the lensing
information encoded in \comic polarization and large-scale structures recovered
from weak lensing galaxy surveys. These quantities allow us to take advantage
of the special features of weak lensing effect on CMB \( B \)-polarization
and of the high (40\%) cross-correlation between the two data sets.
We show that
these objects are robust to filtering effects, have a low intrinsic cosmic variance
(around 8\% for small 100 square degrees surveys) and can be used as an original
constraint on the vacuum energy density.
}
\section{Introduction}
Secondary CMB anisotropies offer new windows to constrain cosmological models. 
Lens effects\cite{TlensEffects} are particularly attractive
since they are expected to be one of the dominant effect. 
Methods to detect the lens effects on CMB\cite{T4pt}
have been proposed recently. Unfortunately all of them suffer from a high sensitivity
to cosmic variance. However, this problem can be solved if one considers
\comic polarization instead of temperature anisotropies. Standard cosmological
models predict that  at small scale, the so called \( B \) component of the polarization can be significant only if CMB-lens couplings are present\cite{B.eq.lent,B2E}. This
feature of CMB polarization will allow us to present tools that enhance the
detection of lens effect in \comic by mixing them with galaxy survey \cite{survlens}.

\section{Lens effects on \comic polarization}

\label{LensEffectSec}

Photons emerging from the last scattering surface are deflected by the large
scale structures of the Universe that are present on the line-of-sights. Therefore
photons observed from apparent direction \( \vec{\alpha } \) must have left
the last scattering surface from a slightly different direction, \( \vec{\alpha }+\vec{\xi }(\vec{\alpha }) \).
Thus, the lensing effect does not produce any polarization nor
rotate the polarization vector, it just moves the apparent
direction of the line of sight\cite{lenspol}; if \(\vec{P}=(Q,U)\) the Stoke variables,  
\begin{equation}
\vec{\hat P}(\vec{\alpha})=\vec{P}(\vec{\alpha}+\vec{\xi}).
\end{equation}
This mechanism alters the geometrical properties of the polarization field, that is to say, changes the \emph{electric} (\( E \)) and \emph{magnetic}
(\( B \)) components of the polarization that reflects its non-local geometrical properties. Indeed, inflationary models, predict that the small scale \( B \) polarization is dominated by lens effect \cite{B.eq.lent,B2E}.
We explicit this point in the weak lensing regime where distortions are small.
At leading order (one can refer to \cite{LeBon}
for description of this calculation) one obtains: 
\begin{equation}
\label{DeltaBdef}
\Delta \hat{B}=-2\epsilon _{ij}\left( \gamma ^{i}\Delta \hat{P}^{j}+\gamma ^{i}_{,k}\hat{P}^{j,k}\right) 
\end{equation}
where we described the lens effect by its convergence field \( \kappa = -1/2\ \xi_{,i}^{i} \) and its shear field \( (\gamma_{1},\gamma_{2})=-1/2\ (\xi_{,x}^{x}-\xi_{,y}^{y}\ ,\ 2\ \xi_{,x}^{y}) \).
The \( \epsilon _{ij} \) (the totally antisymmetric tensor) reflects the geometrical properties of the \( B \) field. It comes in front of two shear-polarization mixing terms. 
One which we will call the \( \bo  \)-term couples the shear with second
derivative of the polarization field. The other one, hereafter the \( \pabo  \)-term,
mixes gradient of the shear and polarization. 

As a consequence, the \( B \) field directly reflects the properties of the
shear maps. Fig. \ref{CompTerm} shows a comparison of relation (\ref{DeltaBdef})
with the exact lensing effect. The agreement is excellent.

\section{Cross-correlating CMB maps and weak lensing surveys}

\label{CrossSec}

With the help of eq.(\ref{DeltaBdef}), one can try to recover lensing information out of \( B \) polarization. Unfortunately, a direct inversion is not possible since it leads to a huge degeneracy in the resulting shear maps \cite{LeBon}.
Another way of deciphering the encoded lensing data will be to cross-correlate
\comic polarization with other lensing information, namely, weak lensing galaxy
surveys\cite{survlens}.

There are strong theoretical motivations to perform this kind of cross-correlations\cite{SperZut,LeBon}. The cross-correlation coefficient between line-of-sight mapped by a photon emerging from last scattering,
\begin{equation}
\label{rDef}
r=\frac{\left\langle \kappa_{\cmbm} \, \kappa _{\galm }\right\rangle }{\sqrt{\left\langle \kappa ^{2}_{\cmbm}\right\rangle \left\langle \kappa _{\galm }^{2}\right\rangle }},
\end{equation}
is around 40\% in standard models and  
accordingly, the correlation between \( B \) polarization and lensing survey
will be significant. Since only the lens effect generates \(B\) field at this scale, we can expect to have a low cosmic variance on the cross-correlation. Moreover, one can assume that systematics and foreground
noises that will hamper each detections will be poorly correlated, so that mixing
the two data sets can be an effective way of enhancing the accuracy of the signal. 

We present here two objects that mix \comic polarization
data with reconstructed shear fields. Looking at eq. (\ref{DeltaBdef}), the
most simple idea is to try to construct \emph{guess} \( B \)-fields, \( b_{\bo} \) and \( b_{\pabo} \), with
local shear instead of the CMB one and to try to correlate
them with our polarization data.
\begin{eqnarray}
b_{\bo } \equiv \epsilon _{ij}\gamma _{\galm }^{i}\Delta \widehat{P}^{j}, \quad
b_{\pabo }\equiv \epsilon _{ij}\partial _{k}\gamma _{\galm }^{i}\partial _{k}\widehat{P}^{j}.
\end{eqnarray}
Then, the amplitude of the cross-correlation between \( \Delta B \)
and \( b_{\bo } \) can easily be estimated. At leading order, we have 
\begin{equation}
\left\langle \Delta \hat{B}\, b_{\bo }(\vec{\alpha })\right\rangle =-\left\langle \Delta E^{2}\right\rangle \left\langle \kappa \kappa _{\galm }\right\rangle, \quad  \left\langle \Delta \hat{B}\, b_{\pabo }(\vec{\alpha })\right\rangle =-\frac{1}{2}\left\langle (\vec{\nabla }E)^{2}\right\rangle \left\langle \vec{\nabla }\kappa \cdot \vec{\nabla }\kappa _{\galm }\right\rangle .
\end{equation}
These results remain valid even when filtering effects are included (see \cite{LeBon} for complete calculation).
\begin{figure}[h]
{\centering \begin{tabular}{cccc}
{\resizebox*{0.22\textwidth}{!}{\includegraphics{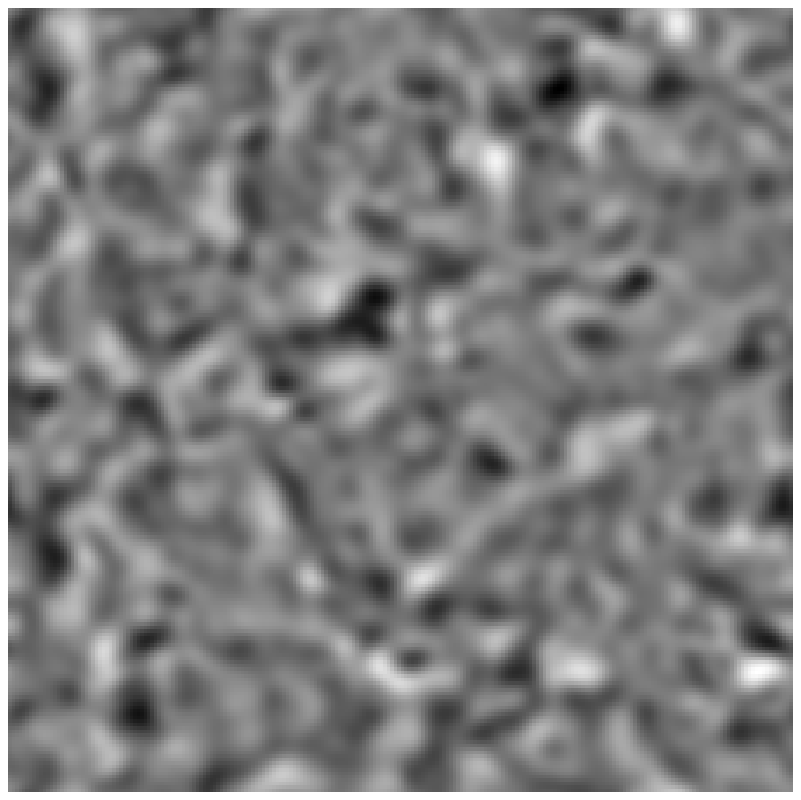}}} &
{\resizebox*{0.22\textwidth}{!}{\includegraphics{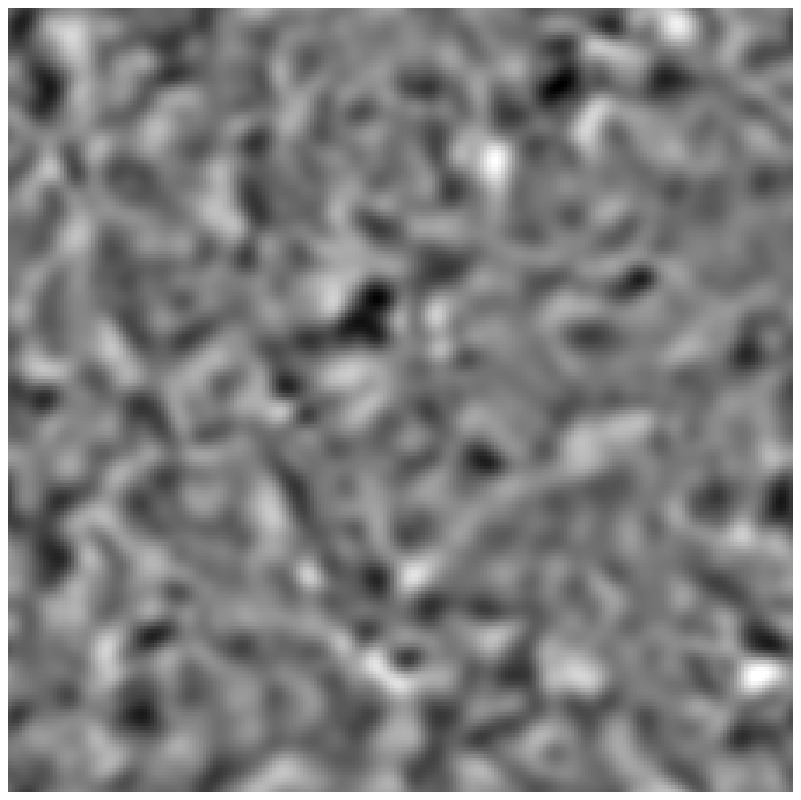}}} &
{\resizebox*{0.22\textwidth}{!}{\includegraphics{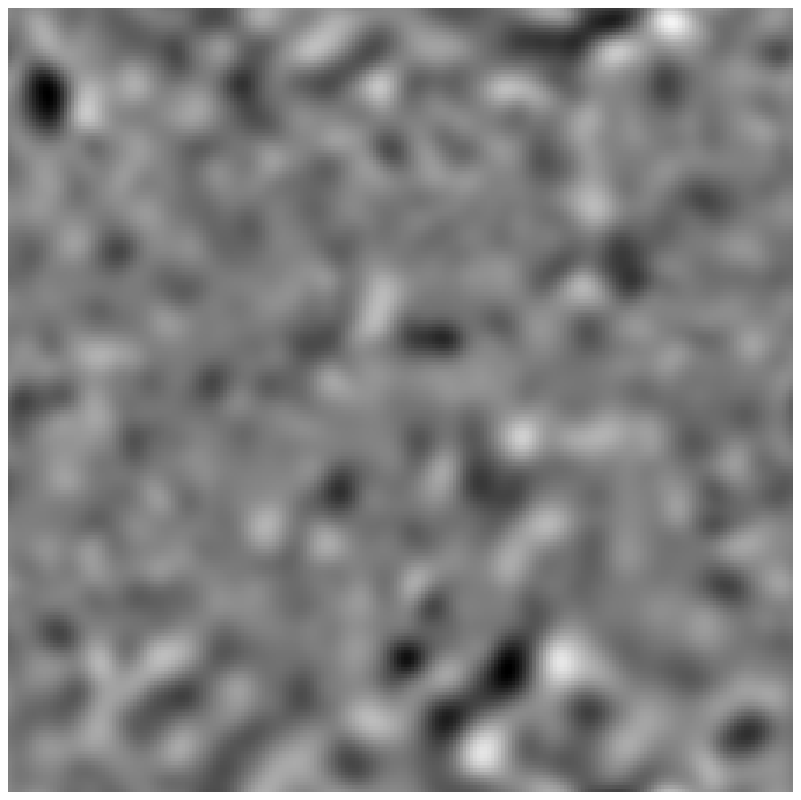}}} &
{\resizebox*{0.22\textwidth}{!}{\includegraphics{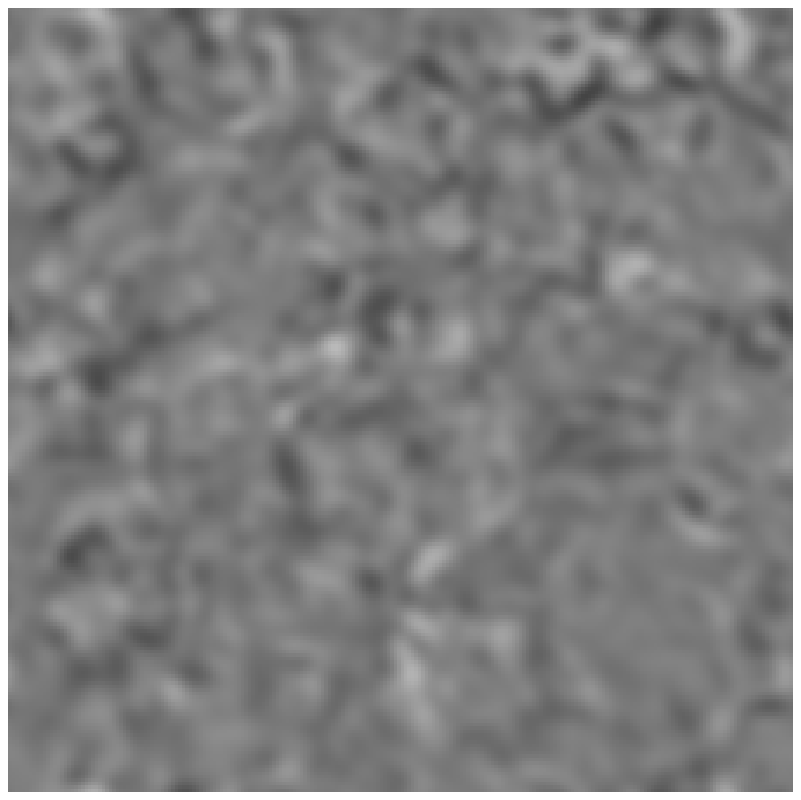}}} 
\end{tabular}\par}

\caption{\label{CompTerm}Numerical simulation results for 4.4 square degrees maps. Right panel is the result of an exact simulation of a \(B\) field. Middle left is the result of the first order approximation. The two right panels are the \(b_{\bo}\) and \(b_{\pabo}\) maps. The convergence fields used here have a cross-correlation coefficient of 0.48. The cross-correlation coefficient between the guess maps and one of the real maps are respectively 0.37 and 0.16.}
\end{figure}

Fig. \ref{CompTerm} shows a numerical illustration of this cross-correlation. 
The similarities between the left and the right maps
are not striking. Yet, under close examination one can recognize individual
patterns shared between the maps. Moreover, computation of correlation coefficient gives significant overlapping up to 40\%.

Using this results, we define two quantities insensitive to the normalization of
\comic and \( \sigma _{8} \) and to filtering effects, which probe the cross-correlation
between two lensing planes:
\begin{eqnarray}
\Cross _{\bo } \equiv   \frac{\left\langle \Delta \hat{B}\, b_{\bo }(\vec{\alpha })\right\rangle }{\left\langle \Delta \hat{E}^{2}\right\rangle \left\langle \kappa _{\galm }^{2}\right\rangle }=-\frac{\left\langle \kappa \kappa _{\galm }\right\rangle }{\left\langle \kappa _{\galm }^{2}\right\rangle }\label{DefObsv1}, \quad
\Cross _{\pabo }  \equiv \frac{\left\langle \Delta \hat{B}\, b_{\pabo }(\vec{\alpha })\right\rangle }{\left\langle (\vec{\nabla }\hat{E})^{2}\right\rangle \left\langle (\vec{\nabla }\kappa _{\galm })^{2}\right\rangle }=-\frac{1}{2}\frac{\left\langle \vec{\nabla }\kappa \cdot \vec{\nabla }\kappa _{\galm }\right\rangle }{\left\langle \vec{\nabla }\kappa _{\galm }^{2}\right\rangle }\label{DefObsv2}
\end{eqnarray}

Previous methods to probe weak lensing in \comic anisotropies \cite{T4pt,SperZut} ran into high cosmic variance problems.
This is not surprising since lens effect which is  not dominant can be masked
by statistical deviations of the primary \comic signal, thus reducing the accuracy of lens detection.
Since \(B\) polarization only emerges in presence of lensing, this last effect should be less important. Indeed, we showed in \cite{LeBon} that \( \Cross_{\bo} \) cosmic variance can be simply estimated in terms of the cosmic variances of the polarization field and of the shear field. 
\begin{eqnarray}
\CosVar (\Cross _{\bo })  = 
\CosVar \left( \left\langle \Delta E^{2}\right\rangle \right) +\left( \frac{1+r^{2}}{2\, r^{2}}\right) \CosVar \left( \left\langle \kappa ^{2}\right\rangle \right) .
\end{eqnarray}
The same kind of equation holds for \( \Cross_{\pabo} \). This expression leads to values for the cosmic variance of \(\Cross_{\bo}\) of less than 8\% for realistic 100 square degrees surveys (see table \ref{CosVar.Bb}).

\begin{table}[h]
\caption{\label{CosVar.Bb}Values of the cosmic variance of \protect\( \Cross _{\natural }\protect \).
The survey size is \protect\( 100\, \textrm{deg}^{2}\protect \). We used the
results of  ray-tracing simulation from \protect \cite{jain} and the \protect\(C_{l}\protect\) given by ``\emph{CMBSlow}''\protect\cite{CMBslow}.
From this estimations, we can expect a cosmic variance for \protect\( \Cross _{\natural }\protect \)of less
than 10\% for realistic scenarii.}
\vspace{0.4cm}
{\centering \begin{tabular}{|c|c|c|c|c|}
\cline{2-5}
\multicolumn{1}{c}{ }&
\multicolumn{2}{|c|}{\( \CosVar \left( \Cross _{\bo }\right)  \) }&
\multicolumn{2}{|c|}{ \( \CosVar \left( \Cross _{\pabo }\right)  \)}\\
\multicolumn{1}{c|}{ }&
\( \Omega _{0}=0.3 \)&
\( \Omega _{0}=1 \)&
\( \Omega _{0}=0.3 \)&
\( \Omega _{0}=1 \)\\
\hline
\( \theta =5',\, \theta _{\galm }=2.5' \) &
6.44\%&
4.77\%&
6.06\%&
4.72\%\\
\( \theta =5',\, \theta _{\galm }=5' \) &
6.58\%&
4.79\%&
4.99\%&
4.23\%\\
\( \theta =10',\, \theta _{\galm }=5' \) &
8.71\%&
6.73\%&
9.49\%&
7.62\%\\
\hline
\end{tabular}\par}
\end{table}

\section{Conclusion - Sensitivity to the cosmic parameters}

We showed that weak lensing effect on the \comic \(B\) polarization can be embedded in a simple, real space, first order expression. This expression can be used to create mathematical objects that compare the lensing effects up to the last scattering surface to the one up to our galaxy surveys probes. We showed that this objects should produce a significant  information, even in realistic (i.e. filtered) situations, with a low intrinsic statistical error. 

\begin{figure}[h]
{\centering \begin{tabular}{cc}
\resizebox*{0.45\textwidth}{!}{\includegraphics{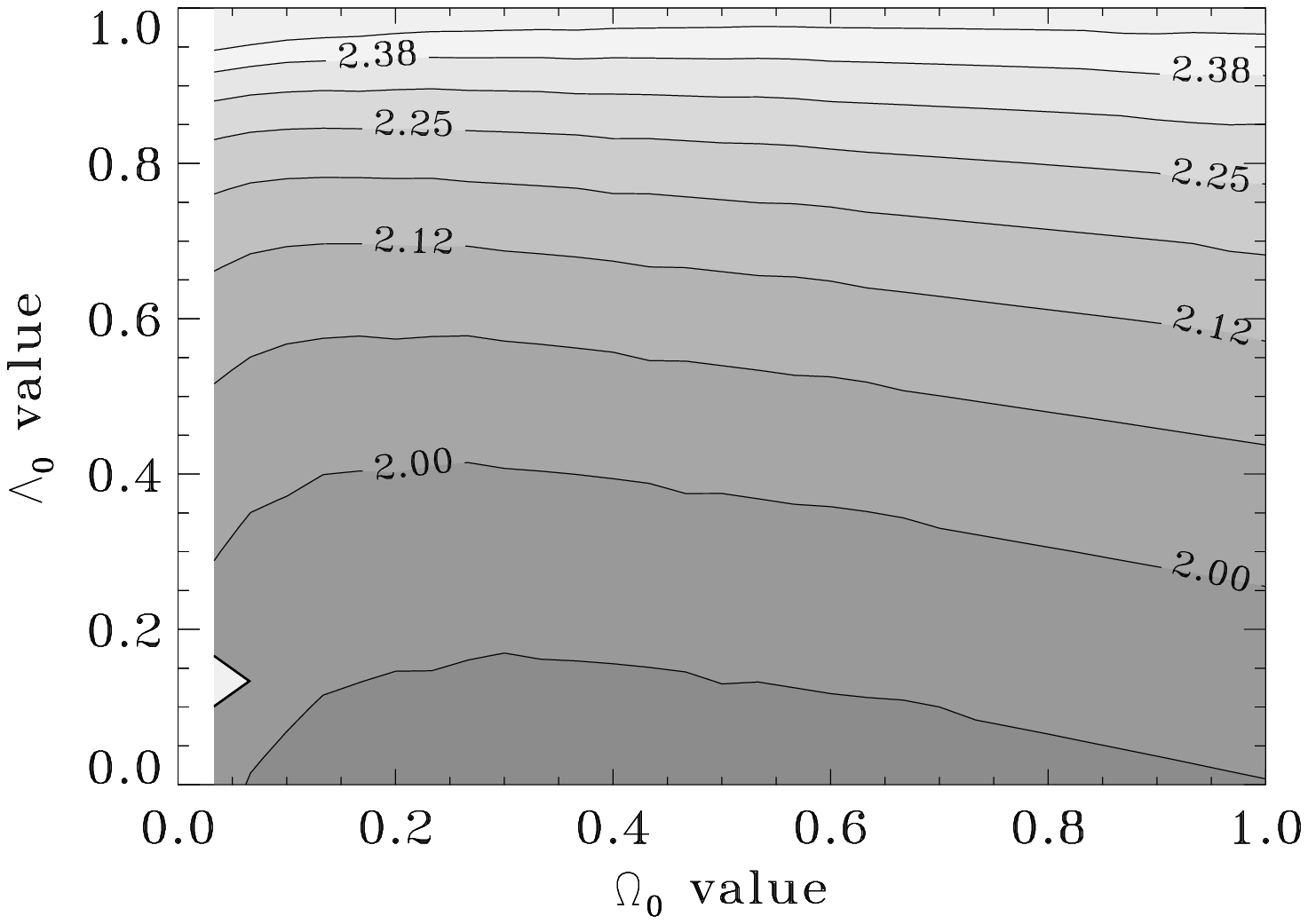}} &
\resizebox*{0.45\textwidth}{!}{\includegraphics{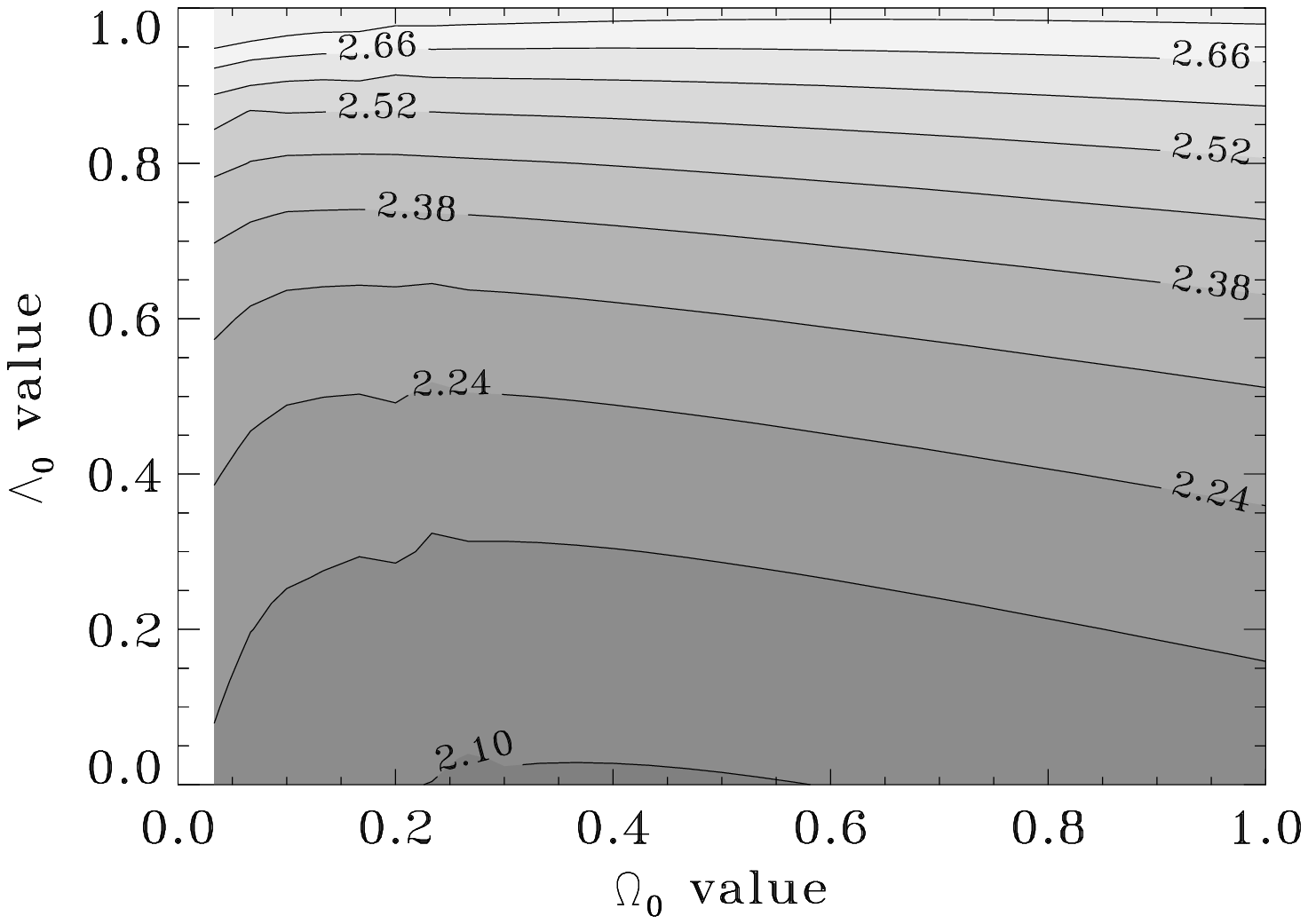}} \\
\end{tabular}\par}

\caption{\label{rparam}\protect\( \Cross_{\bo} \protect \)
(left figure) and \protect\( \Cross_{\pabo} \protect \)
for a CDM model. The filtering beam is 2 arc minutes for all fields. }
\end{figure}

These objects are also expected to be good, unbiased, cosmic parameters tracers. Fig. \ref{rparam} presents their behavior in the \((\Omega_{0},\Lambda)\) plane which exhibit a high sensitivity to the vacuum energy density. This is not surprising, since we are probing the length of the optical bench we are working in, which is rather sensitive to \(\Lambda\) \cite{CalcKappa,LeBon}.

\section*{Acknowledgments}

We thank B. Jain, U. Seljak and S. White for the use of their ray-tracing simulations.
KB and FB thank CITA for hospitality and LvW is thankful to SPhT Saclay for
hospitality. We are all grateful to the TERAPIX data center located at IAP for
providing us computing facilities.

\end{document}